\documentclass[times,sort&compress,3p]{elsarticle}
\journal{Journal of Multivariate Analysis}

\usepackage{amsmath, amssymb, enumerate, amsthm}
\usepackage{lineno,hyperref}

\usepackage{xr-hyper}
\usepackage{bm,amsfonts}
\modulolinenumbers[5]

\usepackage{graphicx,psfrag,epsf,caption,subcaption,booktabs,natbib}
\usepackage{enumitem,tabularx,float}
\usepackage[english]{babel} 
\usepackage[utf8]{inputenc}

\def\spacingset#1{\renewcommand{\baselinestretch}%
{#1}\small\normalsize} \spacingset{1}

\addto\captionsenglish{}

\newtheorem{definition}{Definition}
\theoremstyle{plain}
\newtheorem{theorem}{Theorem}

\newcommand{\numbereqn}{\addtocounter{equation}{1}\tag{\theequation}} 

\begin{document}

\begin{frontmatter}

\title{High-Dimensional Multivariate Posterior Consistency Under Global-Local Shrinkage Priors}

\author[mymainaddress]{Ray Bai\corref{mycorrespondingauthor}}
\cortext[mycorrespondingauthor]{Corresponding author.}
\ead{raybai07@ufl.edu}
\author[mymainaddress]{Malay Ghosh}



\address[mymainaddress]{Department of Statistics, University of Florida, Gainesville, FL 32611, United States}

\begin{abstract}
We consider sparse Bayesian estimation in the classical multivariate linear regression model with $p$ regressors and $q$ response variables. In univariate Bayesian linear regression with a single response $y$, shrinkage priors which can be expressed as scale mixtures of normal densities are popular for obtaining sparse estimates of the coefficients. In this paper, we extend the use of these priors to the multivariate case to estimate a $p \times q$ coefficients matrix $\mathbf{B}$. We derive sufficient conditions for posterior consistency under the Bayesian multivariate linear regression framework and prove that our method achieves posterior consistency even when $p>n$ and even when $p$ grows at nearly exponential rate with the sample size. We derive an efficient Gibbs sampling algorithm and provide the implementation in a comprehensive \textsf{R} package called \texttt{MBSP}. Finally, we demonstrate through simulations and data analysis that our model has excellent finite sample performance.
\end{abstract}


\end{frontmatter}


\section{Introduction} \label{intro}

\subsection{Background} \label{background}
We consider the classical multivariate normal linear regression model,

\begin{align*} \label{Y=XB+E}
\mathbf{Y = XB + E}, \numbereqn
\end{align*}
\noindent
where $\mathbf{Y} = (y_1, \ldots, y_q)$ is an $n \times q$ response matrix of $n$ samples and $q$ continuous response variables, $\mathbf{X}$ is an $n \times p$ matrix of $n$ samples and $p$ covariates, $\mathbf{B} \in \mathbb{R}^{p \times q}$ is the coefficient matrix, and $\mathbf{E} = \left( \varepsilon_1, \ldots, \varepsilon_n \right)^\top$ is an $n \times q$ noise matrix. Under normality, we assume that $\varepsilon_i \overset{\textrm{i.i.d.}}{\sim}  \mathcal{N}_q (\mathbf{0}, \mathbf{\Sigma}), i = 1, \ldots, n$. In other words, each row of $\mathbf{E}$ is identically distributed with mean $\mathbf{0}$ and covariance $\bm{\Sigma}$. Throughout this paper, we also assume that $\mathbf{Y}$ and $\mathbf{X}$ are centered so there is no intercept term in $\mathbf{B}$.

Our focus is on sparse Bayesian estimation and variable selection on the coefficients matrix $\mathbf{B}$ in (\ref{Y=XB+E}).   In practical settings, particularly in high-dimensional settings when $p > n$, it is important not only to provide robust estimates of $\mathbf{B}$, but to choose a subset of regressor variables from the $p$ rows of $\mathbf{B}$ which are good for prediction on the $q$ responses. Although $p$ may be large, the number of predictors that are actually associated with the responses is generally quite small. A parsimonious model also tends to give far better estimation and prediction performance than a dense model, which further motivates the need for sparse estimates of $\mathbf{B}$.

In frequentist approaches to univariate regression, the most commonly used method for inducing sparsity is through imposing regularization penalties on the coefficients of interest. Popular choices of penalty functions include the LASSO \cite{Tibshirani1996} and its many variants, e.g., \cite{SunZhang2012, YuanLin2006, Zou2006, ZouHastie2005}. Many of these penalized regression methods include either an $\ell_1$ penalty function or a combination of an $\ell_1$ and $\ell_2$ penalty to shrink irrelevant predictors or groups of predictors to exactly zero.
 
These $\ell_1$ and $\ell_2$ regularization methods have been naturally extended to the multivariate regression setup where sparsity in the coefficients matrix is desired. For example, \citet{RothmanLevinaZhu2010} utilized an $\ell_1$ penalty on each individual coefficient of $\mathbf{B}$ in (\ref{Y=XB+E}), in addition to an $\ell_1$ penalty on the off-diagonal entries of the covariance matrix to perform joint sparse estimation of $\mathbf{B}$ and $\mathbf{\Sigma}$. \citet{LiNanZhu2015} proposed the multivariate sparse group lasso, which utilizes a combination of a group $\ell_2$ penalty on rows of $\mathbf{B}$ and an $\ell_1$ penalty on the individual coefficients $b_{ij}$ to perform sparse estimation and variable selection at both the group and within-group levels.  \citet{WilmsCroux2017} also consider a model which imposes an $\ell_2$ penalty on the rows of $\mathbf{B}$ to shrink entire rows to zero, while simultaneously estimating the covariance matrix $\mathbf{\Sigma}$. Much of the frequentist literature on sparse estimation of (\ref{Y=XB+E}) has focused on producing robust point estimates of $\mathbf{B}$, rather than on characterizing uncertainty of the estimates. On the other hand, Bayesian methods naturally provide a vehicle for uncertainty quantification through the posterior density. 

In the Bayesian univariate regression model, spike-and-slab priors, introduced by \citet{MitchellBeauchamp1988}, have been a popular choice for inducing sparsity in the coefficients for regression problems. These priors are a mixture density with a point mass at zero used to force some coefficients to zero (the ``spike'') and a continuous density (the ``slab'') to model the nonzero coefficients. Since then, many variants of spike-and-slab have been developed. \citet{GeorgeMcCulloch1993} proposed a stochastic search variable selection (SSVS) method, which places a mixture prior of two normal densities with different variances (one small and one large) on each of the coefficients and which facilitates efficient Gibbs sampling. Recently, \citet{IshwaranRao2005} and \citet{NarisettyHe2014} also used the mixture prior of normals but used rescaling of the variances (dependent upon the sample size $n$) in order to better control the amount of shrinkage for each individual coefficient. In order to perform group estimation and group variable selection, \citet{XuGhosh2015} also introduced the Bayesian group lasso with spike-and-slab  priors (BGL-SS), which is a mixture density with a point mass at a vector $\mathbf{0}_{m_g} \in \mathbb{R}^{m_g}$, where $m_g$ denotes the size of group $g$ and a normal distribution to model the ``slab.''

This two-components mixture approach has been extended to the multivariate framework by several authors \cite{BrownVannucciFearn1998, LiquetBottoloCampanellaRichardsonChadeauHyam2016, LiquetMengersenPettittSutton2017}. In particular, \citet{BrownVannucciFearn1998} and \citet{LiquetBottoloCampanellaRichardsonChadeauHyam2016} first facilitate variable selection by associating each of the $p$ rows of $\mathbf{B}$, $\mathbf{b}_i, 1 \leq i \leq p$, with a $p$-dimensional binary vector $\gamma = (\gamma_1, \ldots, \gamma_p)$, where each entry in $\gamma$ follows a Bernoulli distribution. The selected $\mathbf{b}_i$'s are then estimated by placing a multivariate Zellner $g$-prior (see \citet{zellner1986}) on the sub-matrix of the selected covariates. \citet{LiquetMengersenPettittSutton2017} extend \cite{XuGhosh2015}'s work to the multivariate case with a method called Multivariate Group Selection with Spike and Slab Prior (MBGL-SS). Under MBGL-SS, rows of $\mathbf{B}$ are grouped together and modeled with a prior mixture density with a point mass at $\mathbf{0} \in \mathbb{R}^{m_g q}$ having positive probability (where $m_g$ denotes the size of the $g$th group and $q$ is the number of responses). \citet{LiquetMengersenPettittSutton2017} use the posterior median $\widehat{\mathbf{B}} = (\widehat{b}_{ij})_{p \times q}$ as the estimate for $\mathbf{B}$, so that entire rows are estimated to be exactly zero.

Finally, both frequentist and Bayesian reduced rank regression (RRR) approaches have been developed to tackle the problem of sparse estimation of $\mathbf{B}$ in (\ref{Y=XB+E}). RRR constrains the coefficient matrix $\mathbf{B}$ to be rank-deficient. \citet{ChenHuang2012} proposed a rank-constrained adaptive group lasso approach to recover a low-rank matrix with some rows of $\mathbf{B}$ estimated to be exactly zero. \citet{BuneaSheWegkamp2012} also proposed a joint sparse and low-rank estimation approach and derived its non-asymptotic oracle bounds. The RRR approach was recently adapted to the Bayesian framework by \citet{GohDeyChen2017} and \citet{ZhuKhondkerLuIbrahim2014}. In the Bayesian framework, rank-reducing priors are used to shrink most of the rows and columns in  $\mathbf{B}$ towards $\mathbf{0}_p \in \mathbb{R}^p$ or $\mathbf{0}^\top_q \in \mathbb{R}^q$.

\subsection{Global-local shrinkage priors}
When $p$ is large, (point mass) spike-and-slab priors can face computational problems since they require either searching over $2^p$ possible models. This has led to the creation of a wide number of absolutely continuous shrinkage priors which behave similarly to spike-and-slab priors but which require significantly less computational effort. In univariate regression, these priors can be placed on each individual coefficient $\beta_i, i = 1, \ldots, p,$ and are represented as scale-mixtures of normals:
\begin{equation} \label{globallocal}
\beta_i \overset{\text{ind}}{\sim} \mathcal{N} (0, \tau \xi_i), \hspace{.2cm} \xi_i \sim \pi(\xi_i),
\end{equation}
where $\pi(\xi_i)$ typically follows a heavy-tailed density. These types of priors are known as global-local (GL) shrinkage priors. For GL priors, $\tau$ represents a global parameter that shrinks all coefficients to zero, while $\xi_i$ is a tuning parameter that controls the degree of shrinkage for each individual $\beta_i$. These priors contain significant probability around zero so that most coefficients are shrunk to zero. However, they retain heavy enough tails in order to correctly identify and prevent overshrinkage of the true signals (or non-zero coefficients). This combination of heavy mass around zero and tail robustness makes global-local shrinkage priors especially appealing when inducing sparsity. 

Examples of GL shrinkage priors include the popular horseshoe prior \cite{CarvahoPolsonScott2010} and the Bayesian lasso \cite{ParkCasella2008}. Priors of the type (\ref{globallocal}) have also been considered by numerous authors, including \cite{ArmaganDunsonLee2013, Berger1980, BhadraDattaPolsonWillard2017, GriffinBrown2013, polsonscott2012, Strawderman1971}. \citet{ArmaganClydeDunson2011} noted that a number of these priors utilize a beta prime density as the prior for $\pi(\xi_i)$. This family of generalized beta priors was first studied by \citet{LibbyNovick1982}, and \citet{ArmaganClydeDunson2011} referred to this general class of shrinkage priors as the ``three parameter beta normal'' (TPBN) mixture family. The TPBN family in particular includes the horseshoe, the Strawderman-Berger \cite{Berger1980,Strawderman1971}, and the normal-exponential-gamma (NEG) \cite{GriffinBrown2013} priors. \citet{polsonscott2012} also generalized the beta prime density to the family of hypergeometric inverted beta (HIB) priors. Finally, \citet{ArmaganDunsonLee2013} introduced another general class of priors called the generalized double Pareto (GDP) family. 

These priors have been studied extensively and have been shown to have a number of good theoretical properties. For example, \citet{ArmaganDunsonLeeBajwaStrawn2013} gave sufficient conditions for posterior consistency in univariate linear regression when several well-known shrinkage priors are placed on the coefficients. \citet{GhoshChakrabarti2017}, \citet{VanDerPasKleijnVanderVaart2014}, and \citet{VanDerPasSalmondSchmidtHieber2016} showed that when these priors are used to estimate sparse normal means, the posterior distributions concentrate around the true means at the minimax rate under mild conditions. \citet{VanDerPasSzaboVanDerVaart2017} also obtained minimax-optimal posterior contraction rates for the horseshoe under both empirical Bayes and hierarchical Bayesian choices for the global shrinkage parameter $\tau$ in (\ref{globallocal}).  For the normal means model, the theoretical properties of model selection (including the variable selection method applied in this article) and uncertainty quantification under scale-mixture priors were also recently investigated by \citet{Salomond2017} and \citet{VanDerPasEtAl2017}. Finally, in the context of multiple hypothesis testing, \citet{BhadraDattaPolsonWillard2017}, \citet{DattaGhosh2013}, \citet{GhoshChakrabarti2017}, and \citet{GhoshTangGhoshChakrabarti2016} showed that multiple testing rules induced by these shrinkage priors can achieve optimal Bayes risk in terms of 0-1 symmetric loss (or expected number of misclassified signals). 

\citet{GhoshTangGhoshChakrabarti2016} observed that for a large number of global-local shrinkage priors of the form (\ref{globallocal}), the local parameter  $\xi_i$ has a hyperprior distribution $\pi(\xi_i)$ that can be written as
\begin{align} \label{densityxi}
\pi(\xi_i) = K \xi_i^{-a-1} L(\xi_i),
\end{align}
where $K > 0$ is the constant of proportionality, $a$ is positive real number, and $L$ is a positive measurable, non-constant, slowly varying function over $(0, \infty)$.

\begin{definition} \label{slowlyvaryingdef}
A positive measurable function $L$ defined over $(A, \infty)$, for some $A \geq 0$, is said to be slowly varying (in Karamata's sense) if for every fixed $\alpha > 0, \displaystyle\lim_{x \rightarrow \infty} L(\alpha x)/L(x) = 1$.
\end{definition}
A thorough treatment of functions of this type can be found in the classical text by \citet{Bingham1987}. Table \ref{table:priors} provides a list of several well-known global-local shrinkage priors that fall in the class of priors of the form (\ref{globallocal}), the corresponding density $\pi(\xi_i)$ for $\xi_i$, and the slowly-varying component $L(\xi_i)$ in (\ref{densityxi}). Following \citet{TangXuGhoshGhosh2017}, we refer to these scale-mixture priors as polynomial-tailed priors.

\begin{table}[ttb] 
\centering
\begin{tabular}{ccc}
\hline
Prior & $\pi(\xi_i)/C$ & $L(\xi_i)$  \\
\hline
Student's t & $\xi_i^{-a-1}\exp(-{a}/{\xi_i})$ & $\exp \left( -a/\xi_i \right)$ \\
Horseshoe & $\xi_i^{-1/2}(1+\xi_i)^{-1}$ & $\xi_i^{a+1/2} / (1+\xi_i)$  \\
Horseshoe+ & $ \xi_i^{-1/2}(\xi_i-1)^{-1}\ln(\xi_i)$ & $\xi_i^{a+1/2}(\xi_i - 1)^{-1} \ln (\xi_i)$  \\
NEG & $ \left(1+\xi_i\right)^{-1-a} $ & $\left\{ \xi_i/(1+\xi_i) \right\}^{a+1}$ \\
TPBN & $\xi_i^{u-1}(1+\xi_i)^{-a-u}$ & $\left\{ \xi_i/(1+\xi_i) \right\}^{a+u}$ \\
GDP & $\int_0^\infty \frac{\lambda^2}{2} \exp\left(-\frac{\lambda^2\xi_i}{2}\right)\lambda^{2a-1}\exp(-\eta\lambda)d\lambda$ & $\int_{0}^{\infty} t^{a} \exp(-t - \eta \sqrt{2t/\xi_i}) dt$ \\
HIB & $\xi_i^{u-1}(1+\xi_i)^{-(a+u)}\exp\left( -\frac{s}{1+\xi_i} \right)$ &
$\left\{ \xi_i / (1+\xi_i) \right\}^{a+u}$ \\
& $\times \left( \phi^2+\frac{1-\phi^2}{1+\xi_i}\right)^{-1}$ &  $\times   \exp \left( - \frac{s}{1 + \xi_i} \right) \left( \phi^2 + \frac{1- \phi^2}{1 + \xi_i} \right)^{-1}$ \\
\hline
\end{tabular} 
\caption{Polynomial-tailed priors, their respective prior densities for $\pi(\xi_i)$ up to normalizing constant $C$, and the slowly-varying component $L(\xi_i)$.  }\label{table:priors}
\end{table} 

Although polynomial-tailed priors have been studied extensively in univariate regression, their potential utility for multivariate analysis seems to have been largely overlooked. In this paper, we introduce a new Bayesian approach for estimating the unknown $p \times q$ coefficient matrix $\mathbf{B}$ in (\ref{Y=XB+E}) using polynomial-tailed priors. We call our method the Multivariate Bayesian model with Shrinkage Priors (MBSP). 

While there have been many methodological developments for Bayesian multivariate linear regression, theoretical results in this domain have not kept pace with applications. There appears to be very little theoretical justification for adopting Bayesian methodology in multivariate regression. In this article, we take a step towards resolving this gap by providing sufficient conditions under which Bayesian multivariate linear regression models can obtain posterior consistency. To our knowledge, our paper is the first one to give general conditions for posterior consistency under model (\ref{Y=XB+E}) when $p>n$ and when $p$ grows at nearly exponential rate with sample size $n$. We further illustrate that our method based on polynomial-tailed priors achieves strong posterior consistency in both low-dimensional and ultrahigh-dimensional settings.

The rest of our paper is organized as follows. In Section \ref{mbsp}, we introduce the MBSP model and provide some insight into how it facilitates sparse estimation and variable selection. In Section \ref{consistency}, we present sufficient conditions for our model to achieve posterior consistency in both the cases where $p$ grows slower than $n$ and the case when $p$ grows at nearly exponential rate with $n$. In Section \ref{Implementation}, we show how to implement MBSP using the TPBN family of priors as a special case and how to utilize our method for variable selection. Efficient Gibbs sampling and computational complexity considerations are also discussed. In Section \ref{SimulationsAndDataAnalysis}, we illustrate our method's finite sample performance through simulations and analysis of a real data set. Finally, in Section \ref{Conclusion}, we discuss some directions for future research.

\section{Multivariate Bayesian Model with Shrinkage Priors (MBSP)} \label{mbsp}

\subsection{Preliminary Notation and Definitions}
We first introduce the following notation and definitions.

\begin{definition} \label{matrixnormaldist}
A random matrix $\mathbf{Y}$ is said to have the \textit{matrix-normal density} if $\mathbf{Y}$ has the density function (on the space $\mathbb{R}^{a \times b}$):
\begin{align} \label{matrix-normal}
f(\mathbf{Y}) = \frac{| \mathbf{U} |^{-b/2} | \mathbf{V} |^{-a/2}}{(2 \pi)^{ab/2}} e^{ - \frac{1}{2} \textrm{tr} [ \mathbf{U}^{-1} ( \mathbf{Y} - \mathbf{M} ) \mathbf{V}^{-1} ( \mathbf{Y} - \mathbf{M})^\top ]},  
\end{align} 
where $\mathbf{M} \in \mathbb{R}^{a \times b}$, and $\mathbf{U}$ and $\mathbf{V}$ are positive definite matrices of dimension $a \times a$ and $b \times b$ respectively. If $\mathbf{Y}$ is distributed as a matrix-normal distribution with pdf given in (\ref{matrix-normal}), we write $\mathbf{Y} \sim \mathcal{MN}_{a \times b} (\textbf{M, U, V})$. 
\end{definition}

\begin{definition}
The matrix $\mathbf{O} \in \mathbb{R}^{a \times b}$ denotes the $a \times b$ matrix with all zero entries. 
\end{definition}

\subsection{MBSP Model} \label{modeldescr}
Our multivariate Bayesian model formulation for model (\ref{Y=XB+E}) with shrinkage priors (henceforth referred to as MBSP) is as follows:

\begin{align*}\label{mbspmodel}
\begin{array}{rl}
\mathbf{Y| X, B, \Sigma} 
	&\sim \mathcal{MN}_{n \times q} (\mathbf{XB}, \mathbf{I}_n, \mathbf{\Sigma}),  \numbereqn \\ 
	\mathbf{B} | \xi_1, \ldots, \xi_p, \mathbf{\Sigma} & \sim \mathcal{MN}_{p \times q} (\mathbf{O}, \tau \: \text{diag}(\xi_1, \ldots, \xi_p), \mathbf{\Sigma}),   \\ 
	\xi_i & \overset{\text{ind}}{\sim} \pi(\xi_i), i= 1, \ldots, p,   
\end{array}
 \end{align*} 
where $\pi(\xi_i)$ is a polynomial-tailed prior density of the form (\ref{densityxi}). 

\subsection{Handling Sparsity}
In this section, we illustrate how the MBSP model induces sparsity. First note that in  (\ref{mbspmodel}), an alternative way of writing the density $\mathbf{Y} | \mathbf{X}, \mathbf{B}, \mathbf{\Sigma}$ is
\begin{equation}
\mathbf{Y} | \mathbf{X}, \mathbf{B}, \mathbf{\Sigma} \propto | \mathbf{\Sigma} |^{-nq/2} \exp \left\{ -\frac{1}{2} \displaystyle \sum_{i=1}^n \left(\mathbf{y}_i - \displaystyle\sum_{j=1}^p x_{ij} \mathbf{b}_j \right)^\top \mathbf{\Sigma}^{-1} \left( \mathbf{y}_i - \displaystyle\sum_{j=1}^p x_{ij} \mathbf{b}_j \right) \right\}, \label{altpdf}
\end{equation}
where $\mathbf{b}_j$ denotes the $j$th row of $\mathbf{B}$.

Following from (\ref{altpdf}), we see that under (\ref{mbspmodel}) and known $\mathbf{\Sigma}$, the joint prior density $\pi(\mathbf{B}, \xi_1, \ldots, \xi_p)$ is 
\begin{align}\label{Bandxis}
\pi(\mathbf{B}, \xi_1, \ldots, \xi_p) \propto \displaystyle \prod_{j=1}^p \xi_j^{-q/2} e^{-\frac{1}{2 \xi_j} || \mathbf{b}_{j} ( \tau \mathbf{\Sigma})^{-1/2} ||_2^2} \pi(\xi_j),
\end{align}
\noindent
where $|| \cdot ||_2$ denotes the $\ell_2$ vector norm. Since the $p$ rows of $\mathbf{B}$ are independent, we see from (\ref{Bandxis}) that this choice of prior induces sparsity on the rows of $\mathbf{B}$, while also accounting for the covariance structure of the $q$ responses. This ultimately facilitates sparse estimation of $\mathbf{B}$ as a whole and variable selection from the $p$ regressors. 

For example, if $\pi( \xi_j) \overset{\text{ind}}{\sim} \mathcal{IG} (\alpha_j, \frac{\gamma_j}{2})$ (where $\mathcal{IG}$ denotes the inverse-gamma density), then the marginal density for $\mathbf{B}$ (after integrating out the $\xi_j$'s) is proportional to
\begin{align}\label{multit}
\displaystyle\prod_{j=1}^p \left( || \mathbf{b}_{j} (\tau \mathbf{\Sigma})^{-1/2} ||_2^2 + \gamma_j \right)^{- (\alpha_j + \frac{q}{2}) },
\end{align}
which corresponds to a multivariate Student's $t$ density.

On the other hand, if $\pi(\xi_j) \propto \xi_j^{q/2-1} (1+ \xi_j)^{-1}$, then the joint density in \label{pi(Bandxis)} is proportional to
\begin{align}\label{multihs}
\displaystyle\prod_{j=1}^p \xi_j^{-1} (1+\xi_j)^{-1} e^{ -\frac{1}{2 \xi_j} || \mathbf{b}_{j} (\tau \mathbf{\Sigma})^{-1/2} ||_2^2  },
\end{align}
and integrating out the $\xi_j$'s gives a multivariate horseshoe density function.

As examples (\ref{multit}) and (\ref{multihs}) demonstrate, our model allows us to obtain sparse estimates of $\mathbf{B}$ by inducing row-wise sparsity in $\mathbf{B}$ with a matrix-normal scale mixture using global-local shrinkage priors. This row-wise sparsity also facilitates variable selection from the $p$ variables.

\section{Posterior Consistency of MBSP} \label{consistency}

\subsection{Notation} \label{notation}
We first introduce some notation that will be used throughout the paper. For any two sequences of positive real numbers $\{ a_n \}$ and $\{ b_n \}$ with $b_n \neq 0$, we write $a_n = O(b_n)$ if $ \left|  a_n/b_n \right| \leq M$ for all $n$, for some positive real number $M$ independent of $n$, and $a_n = o(b_n)$ to denote $\lim_{n \rightarrow \infty} a_n/ b_n = 0$. Therefore, $a_n = o(1)$ if $\lim_{n \rightarrow \infty} a_n = 0$.

For a vector $v \in \mathbb{R}^n$, $|| v ||_2 := \sqrt{ \sum_{i=1}^n v_i^2}$ denote the $\ell_2$ norm. For a matrix $\mathbf{A} \in \mathbb{R}^{a \times b}$ with entries $a_{ij}$, $|| \mathbf{A} ||_F := \sqrt{\textrm{tr} (\mathbf{A}^T \mathbf{A})} = \sqrt{\sum_{i=1}^{a} \sum_{j=1}^b a_{ij}^2}$ denotes the Frobenius norm of $\mathbf{A}$. For a symmetric matrix $\mathbf{A}$, we denote its minimum and maximum eigenvalues by $\lambda_{\text{min}} (\mathbf{A})$ and $\lambda_{\text{max}}(\mathbf{A})$ respectively. Finally, for an arbitrary set $\mathcal{A}$, we denote its cardinality by $| \mathcal{A} |$.

\subsection{Definition of Posterior Consistency} \label{defconsistency}
For this section, we denote the number of predictors by $p_n$ to emphasize that $p$ depends on $n$ and is allowed to grow with $n$. Suppose that the true model is
\begin{equation} \label{B0truth}
\mathbf{Y}_n = \mathbf{X}_n \mathbf{B}_{0n} + \mathbf{E}_n,
\end{equation}
where $\mathbf{Y}_n := (\mathbf{Y}_{n,1}, ..., \mathbf{Y}_{n,q})$ and $\mathbf{E}_n \sim \mathcal{MN}_{n \times q} (\mathbf{O}, \mathbf{I}_n, \mathbf{\Sigma})$.  For convenience, we denote $\mathbf{B}_{0n}$ as $\mathbf{B}_0$ going forward, noting $\mathbf{B}_0$ depends on $p_n$ (and therefore on $n$).  

Let $\{ \mathbf{B}_{0} \}_{n \geq 1}$ be the sequence of true coefficient matrices, and let $\mathbb{P}_0$ denote the distribution of $\{ \mathbf{Y}_n \}_{n \geq 1}$ under (\ref{B0truth}).  Let $\{ \pi_n (\mathbf{B}_n) \}_{n \geq 1}$ and $\{ \pi_n (\mathbf{B}_n | \mathbf{Y}_n) \}_{n \geq 1}$ denote the sequences of prior and posterior densities for coefficients matrix $\mathbf{B}_n$. Analogously, let $\{ \Pi_n (\mathbf{B}_n ) \}_{n \geq 1}$ and $\{ \Pi_n (\mathbf{B}_n | \mathbf{Y}_n ) \}_{n \geq 1}$ denote the sequences of prior and posterior distributions.  In order to achieve consistent estimation of $\mathbf{B}_0  (\equiv \mathbf{B}_{0n})$, the posterior probability that $\mathbf{B}_n$ lies in a $\varepsilon$-neighborhood of $\mathbf{B}_{0}$ should converge to 1 almost surely with respect to $\mathbb{P}_0$ measure as $n \rightarrow \infty$. We therefore define strong posterior consistency as follows:

\begin{definition} \label{consistencydef}
\textbf{(posterior consistency)} Let $\mathcal{B}_n = \{ \mathbf{B}_n : || \mathbf{B}_n - \mathbf{B}_0 ||_F > \varepsilon \}$, where $\varepsilon > 0$. The sequence of posterior distributions of $\mathbf{B}_n$ under prior $\pi_n (\mathbf{B}_n)$ is said to be  \textit{strongly} consistent under (\ref{B0truth}) if, for any $\varepsilon > 0$,
\begin{equation*}
 \Pi_n (\mathcal{B}_n | Y_n ) = \Pi_n ( || \mathbf{B}_n - \mathbf{B}_0 ||_F > \varepsilon | Y_n ) \rightarrow 0 \textrm{ a.s. } \mathbb{P}_0 \textrm{ as } n \rightarrow \infty.
\end{equation*}  
\end{definition}

Using Definition \ref{consistencydef}, we now state two general theorems and a corollary that provide general conditions under which priors on $\mathbf{B}$ (not just the MBSP model) may achieve strong posterior consistency in both low-dimensional and ultrahigh-dimensional settings.

\subsection{Sufficient Conditions for Posterior Consistency} \label{generalconsistency}
For our theoretical investigation, we assume $\mathbf{\Sigma}$ to be fixed and known and dimension of the response variables $q$ to be fixed. In practice, $\mathbf{\Sigma}$ is typically unknown, and one can estimate it from the data. In Section \ref{Implementation}, we present a fully Bayesian implementation of MBSP by placing an appropriate inverse-Wishart prior on $\mathbf{\Sigma}$. 

Theorem \ref{Th:1} applies to the case where the number of predictors $p_n$ diverges to $\infty$ at a rate slower than $n$ as $n \rightarrow \infty$, while Theorem \ref{Th:2} applies to the case where $p_n$ grows to $\infty$ at a faster rate than $n$ as $n \rightarrow \infty$. To handle these two cases, we require different sets of regularity assumptions. Proofs for both theorems are shown in Section 1 of the supplementary materials (see \ref{App:A}).

\subsubsection{Low-Dimensional Case}
We first impose the following regularity conditions which are all standard ones used in the literature and relatively mild (see, for example, \citet{ArmaganDunsonLeeBajwaStrawn2013}). In particular, Assumption \ref{As:A2} ensures that the design matrix $\mathbf{X}_n^T \mathbf{X}_n$ is positive definite for all $n$ and that $\mathbf{B}_0$ is estimable. 

\subsubsection*{Regularity Conditions }
\begin{enumerate}[label=(A\arabic*)]
\item
$p_n = o(n)$ and $p_n \leq n$ for all $n \geq 1$. \label{As:A1}
\item
There exist constants $c_1, c_2$ so that
\begin{equation*}
0 < c_1 < \lim \displaystyle\inf_{n \rightarrow \infty} \lambda_{\min} \left( \frac{\mathbf{X}_n^\top \mathbf{X}_n}{n} \right) \leq \lim \displaystyle\sup_{n \rightarrow \infty} \lambda_{\max} \left( \frac{\mathbf{X}_n^\top \mathbf{X}_n}{n} \right) < c_2 < \infty.
\end{equation*} \label{As:A2}
\vspace{-.5cm}
\item
There exist constants $d_1$ and $d_2$ so that 
\begin{equation*}
0 < d_1 < \lambda_{\textrm{min}} (\mathbf{\Sigma}) \leq \lambda_{\textrm{max}} (\mathbf{\Sigma}) < d_2 < \infty.
\end{equation*} \label{As:A3}
\end{enumerate}
\vspace{-.5cm}
Using these conditions, we are able to attain a very simple sufficient condition for strong posterior consistency under (\ref{B0truth}), as defined in Definition \ref{consistencydef}, which we state in the next theorem.

\begin{theorem}
\label{Th:1}
Assume that conditions \ref{As:A1}-\ref{As:A3} hold. Then the posterior of $\mathbf{B}_n$ under any prior $\pi_n (\mathbf{B}_n)$ is strongly consistent under (\ref{B0truth}), i.e. for $\mathcal{B}_n = \{ \mathbf{B}_n : || \mathbf{B}_n - \mathbf{B}_0 ||_F > \varepsilon \}$ and any arbitrary $\varepsilon > 0$,
\begin{equation*}
\Pi_n (\mathcal{B}_n | \mathbf{Y}_n) \rightarrow 0 \textrm{ a.s. } \mathbb{P}_0 \textrm{ as } n \rightarrow \infty
\end{equation*}
if 
\begin{align} \label{posteriorBn}
\Pi_n \left( \mathbf{B}_n: || \mathbf{B}_n - \mathbf{B}_0 ||_F < \frac{\Delta}{n^{\rho /2} } \right) > \exp (-kn)
\end{align}
for all $0 < \Delta < \frac{\varepsilon^2 c_1 d_1^{1/2}}{48 c_2^{1/2} d_2}$ and $0 < k  < \frac{ \varepsilon^2 c_1}{32 d_2} - \frac{3 \Delta c_2^{1/2}}{2 d_1^{1/2}}$, where $\rho > 0$.
\end{theorem}
Condition in (\ref{posteriorBn}) in Theorem \ref{Th:1} states that as long as the prior distribution for $\mathbf{B}_n$ captures $\mathbf{B}_0$ inside a ball of radius $\Delta / n^{\rho/2}$ with sufficiently high probability for large $n$, the posterior of $\mathbf{B}_n$ will be strongly consistent. 

\subsubsection{Ultrahigh Dimensional Case}
To achieve posterior consistency when $p_n \gg n$ and $p_n \geq O(n)$, we require additional restrictions on the eigenstructure of $\mathbf{X}_n$ and an additional assumption on the size of the true model. Working under the assumption of sparsity, we assume that the true model (\ref{B0truth}) contains only a few nonzero predictors. That is, most of the rows of $\mathbf{B}_0$ should contain only zero entries.We denote $S^* \subset \{1, 2, ..., p_n \}$ as the set of indices of the rows of $\mathbf{B}_0$ with at least one nonzero entry and let $s^* = |S^*|$ be the size of $S^*$. We need the following regularity conditions.

\subsubsection*{Regularity Conditions}
\begin{enumerate}[label=(B\arabic*)]
\item
$p_n > n$ for all $n \geq 1$, and $\ln (p_n) = O(n^d)$ for some $0 < d < 1$. \label{As:B1}
\item
The rank of $\mathbf{X}_n$ is $n$. \label{As:B2}
\item
Let $\mathcal{J}$ denote a set of indices, where $\mathcal{J} \subset \{1, ..., p_n \}$ such that $| \mathcal{J} | \leq n$. Let $\mathbf{X}_{\mathcal{J}}$ denote the submatrix of $\mathbf{X}$ that contains the columns with indices in $\mathcal{J}$. For any such set $\mathcal{J}$, there exists a finite constant $\widetilde{c}_1 (> 0)$ so that 
\begin{equation*}
\lim \inf_{n \rightarrow \infty} \lambda_{\textrm{min}} \left( \frac{ \mathbf{X}_{\mathcal{J}}^\top \mathbf{X}_{\mathcal{J}}}{n} \right) \geq \widetilde{c}_1.
\end{equation*} \label{As:B3}
\vspace{-.5cm}
\item
There is finite constant $\widetilde{c}_2 (>0)$ so that 
\begin{equation*}
\lim \sup_{n \rightarrow \infty} \lambda_{\textrm{max}} \left( \frac{\mathbf{X}_n^\top \mathbf{X}_n}{n} \right) < \widetilde{c}_2.
\end{equation*} \label{As:B4}
\vspace{-.5cm}
\item
There exist constants $d_1$ and $d_2$ so that 
\begin{equation*}
0 < d_1 < \lambda_{\textrm{min}} (\mathbf{\Sigma}) \leq \lambda_{\textrm{max}} (\mathbf{\Sigma}) < d_2 < \infty.
\end{equation*} \label{As:B5}
\vspace{-.5cm}
\item
$S^*$ is nonempty for all $n \geq 1$, and $s^* = o(n / \ln (p_n)).$ \label{As:B6}
\end{enumerate}

Condition \ref{As:B1} allows the number of predictors $p_n$ to grow at nearly exponential rate. In particular, $p_n$ may grow at a rate of $e^{n^d}$, where $0 < d < 1$. In the high-dimensional literature, it is a standard assumption that $\ln (p_n) = o(n)$. Condition \ref{As:B3} assumes that for any submatrix of $\mathbf{X}_n$ that is full rank, its minimum singular value is bounded below by $n \widetilde{c}_1$. This condition is needed to overcome potential identifiability issues, since trivially, the smallest singular value of $\mathbf{X}_n$ is zero. \ref{As:B4} imposes a supremum on the maximum singular value of $\mathbf{X}_n$, which poses no issue. Finally, Condition \ref{As:B6} allows the true model size to grow with $n$ but at a rate slower than $n / \ln (p_n)$. \ref{As:B6} is a standard condition that has been used to establish estimation consistency when $p_n$ grows at nearly exponential rate with $n$ for frequentist point estimators, such as the Dantzig estimator \cite{CandesTao2007}, the scaled LASSO \cite{SunZhang2012}, and the LASSO \cite{Tibshirani1996}. In ultrahigh-dimensional problems, it is generally agreed that $s^*$ must be small relative to both $p$ and $n$ in order to attain estimation consistency and minimax convergence rates, and hence, this restriction on the growth rate of $s^*$.

Under these regularity conditions, we are able to attain a simple sufficient condition for posterior consistency under (\ref{B0truth}) even when $p_n$ grows faster than $n$. Theorem \ref{Th:2} gives the sufficient condition for strong consistency.

\begin{theorem}
\label{Th:2}
Assume that conditions \ref{As:B1}-\ref{As:B6} hold. Then the posterior of $\mathbf{B}_n$ under any prior $\pi_n (\mathbf{B}_n)$ is strongly consistent under (\ref{B0truth}), i.e. for $\mathcal{B}_n = \{ \mathbf{B}_n : || \mathbf{B}_n - \mathbf{B}_0 ||_F > \varepsilon \}$ and any arbitrary $\varepsilon > 0$,
\begin{equation*}
\Pi_n (\mathcal{B}_n | \mathbf{Y}_n) \rightarrow 0 \textrm{ a.s. } \mathbb{P}_0 \textrm{ as } n \rightarrow \infty
\end{equation*}
if 
\vspace{-.3cm}
\begin{align} \label{posteriorBn2}
\Pi_n \left( \mathbf{B}_n: || \mathbf{B}_n - \mathbf{B}_0 ||_F < \frac{\widetilde{\Delta}}{n^{\rho /2} } \right) > \exp (-kn)
\end{align}
for all $0 < \widetilde{\Delta} < \frac{\varepsilon^2 \widetilde{c}_1 d_1^{1/2}}{48 \widetilde{c}_2^{1/2} d_2}$ and $0 < k  < \frac{ \varepsilon^2 \widetilde{c}_1}{32 d_2} - \frac{3 \widetilde{\Delta} \widetilde{c}_2^{1/2}}{2 d_1^{1/2}}$, where $\rho > 0$.
\end{theorem}

Similar to (\ref{posteriorBn}) in Theorem \ref{Th:1}, condition (\ref{posteriorBn2}) in Theorem \ref{Th:2} states that as long as the prior distribution for $\mathbf{B}_n$ captures $\mathbf{B}_0$ inside a ball of radius $\widetilde{\Delta} / n^{\rho/2}$ with sufficiently high probability for large $n$, the posterior of $\mathbf{B}_n$ will be strongly consistent. To our knowledge, our paper is the first one in the literature to address the issue of ultra high-dimensional consistency in Bayesian multivariate linear regression. There has been very little theoretical investigation done in the framework of Bayesian multivariate regression, and our paper takes a step towards narrowing this gap.

Now that we have provided simple sufficient conditions for posterior consistency in Theorems \ref{Th:1} and \ref{Th:2}, we are ready to state our main theorems which demonstrate the power of the MBSP model (\ref{mbspmodel}) under polynomial-tailed hyperpriors (\ref{densityxi}).

\subsection{Sufficient Conditions for Posterior Consistency of MBSP} \label{mbspconsistent}
We now establish posterior consistency under the MBSP model (\ref{mbspmodel}), assuming that $\mathbf{\Sigma}$ is fixed and known, $q$ is fixed, and that $\tau = \tau_n$ is a tuning parameter that depends on $n$.

As in Section \ref{generalconsistency}, we assume that most of the rows of $\mathbf{B}_0$ are zero, i.e. that the true model $S \subset \{1, ..., p_n \}$ is small relative to the total number of predictors. As before, we consider the cases where $p_n = o(n)$ and $p_n \geq O(n)$ separately. We also require the following regularity assumptions which turn out to be sufficient for both the low-dimensional and ultra high-dimensional cases. Here, $b_{jk}^0$ denotes an entry in $\mathbf{B}_0$.

\subsubsection*{Regularity Conditions}
\begin{enumerate}[label=(C\arabic*)]
\item
For the slowly varying function $L(t)$ in the priors for $\xi_i, 1 \leq i \leq p$, in (\ref{densityxi}), $\lim_{t \rightarrow \infty} L(t) \in (0, \infty)$. That is, there exists $c_0 (> 0)$ such that $L(t) \geq c_0$ for all $t \geq t_0$, for some $t_0$ which depends on both $L$ and $c_0$. \label{As:C1}
\item
There exists $M > 0$ so that $\sup_{j,k} | b_{jk}^{0} | \leq M < \infty$ for all $n$, i.e. the maximum entry in $\mathbf{B}_0$ is uniformly bounded above in absolute value. \label{As:C2}
\item
$0 < \tau_n < 1$ for all $n$, and $\tau_n = o (p_n^{-1} n^{-\rho} )$ for $\rho > 0$. \label{As:C3} 
\end{enumerate}
\noindent \textbf{Remark 1.} Condition \ref{As:C1} is a very mild condition which ensures that $L(\cdot)$ is slow-varying. \citet{GhoshTangGhoshChakrabarti2016} established that \ref{As:C1} holds for $L(\cdot)$ in the TPBN  priors ($L(\xi_i) = (1+\xi_i)^{-(\alpha+\beta)}$) and the GDP priors ($L(\xi_i) = 2^{-\frac{\alpha}{2} -1} \int_0^{\infty} e^{-\beta \sqrt{2u/ \xi_i}}$ $ e^{-u}$ $u^{(\frac{\alpha}{2}+1)-1}du$). The TPBN family in particular includes many well-known one-group shrinkage priors, such as the horseshoe prior ($\alpha =0.5, \beta = 0.5$), the Strawderman-Berger prior ($\alpha = 1, \beta = 0.5$), and the normal-exponential-gamma prior ($\alpha = 1, \beta > 0$). As remarked by \citet{GhoshChakrabarti2017}, one easily verifies that Assumption \ref{As:C1} also holds for the inverse-gamma priors ($\pi(\xi_i) \propto \xi_i^{-\alpha-1} e^{-b/ \xi_i }$) and the half-t priors ($\pi(\xi_i) \propto (1 + \xi / \nu)^{-(\nu + 1)/2}$).
\vspace{.5cm}

\noindent \textbf{Remark 2.} Condition \ref{As:C2} is a mild condition that bounds the entries of $\mathbf{B}_0$ in absolute value for all $n$, while \ref{As:C3} specifies an appropriate rate of decay for $\tau_n$. It is possible that the upper bound on the rate for $\tau_n$ can be loosened for individual GL priors. However, since we wish to encompass all possible priors of the form (\ref{densityxi}), we provide a general rate that works for all the polynomial-tailed priors considered in this paper.
\vspace{.5cm}

We are now ready to state our main theorem for posterior consistency of the MBSP model. The proof for Theorem \ref{Th:3} can be found in Section 2 of the supplementary materials (see \ref{App:A}).

\begin{theorem}[\textbf{low-dimensional case}]
\label{Th:3}
Suppose that we have the MBSP model (\ref{mbspmodel}) with hyperpriors (\ref{densityxi}). Provided that Assumptions \ref{As:A1}-\ref{As:A3} and \ref{As:C1}-\ref{As:C3} hold, our model achieves strong posterior consistency. That is, for any $\varepsilon > 0$,
\begin{equation*}
\Pi_n ( \mathbf{B}_n: || \mathbf{B}_n - \mathbf{B}_0 ||_F > \varepsilon | \mathbf{Y}_n ) \rightarrow 0 \hspace{.2cm} \textrm{ a.s. } \mathbb{P}_0 \textrm{ as } n \rightarrow \infty.
\end{equation*} 
\end{theorem}
Theorem \ref{Th:3} establishes posterior consistency for the MBSP model only when $p_n = o(n)$. We also note that in the low-dimensional setting where $p_n = o(n)$, we place \textit{no} restrictions on the growth on the number of nonzero predictors in the true model relative to sample size $n$. This contrasts with a previous result by \citet{ArmaganDunsonLeeBajwaStrawn2013}, who required that the number of true nonzero covariates grow slower than $n / \ln(n)$.

In the ultra high-dimensional case where $p_n \geq O(n)$, we can still achieve posterior consistency under the MBSP model, with additional mild restrictions on the design matrix $\mathbf{X}_n$ and on the size of the true model. Theorem \ref{Th:4} deals with the ultra high-dimensional scenario. The proof for Theorem \ref{Th:4} can be found in Section 2 of the supplementary materials (see \ref{App:A}).

\begin{theorem}[\textbf{ultra high-dimensional case}]
\label{Th:4}
Suppose that we have the MBSP model (\ref{mbspmodel}) with hyperpriors (\ref{densityxi}). Provided that Assumptions \ref{As:B1}-\ref{As:B6} and \ref{As:C1}-\ref{As:C3} hold, our model achieves strong posterior consistency. That is, for any $\varepsilon > 0$,
\begin{equation*}
\Pi_n ( \mathbf{B}_n: || \mathbf{B}_n - \mathbf{B}_0 ||_F > \varepsilon | \mathbf{Y}_n ) \rightarrow 0 \hspace{.2cm} \textrm{ a.s. } \mathbb{P}_0 \textrm{ as } n \rightarrow \infty.
\end{equation*} 
\end{theorem}

Interestingly enough, to ensure posterior consistency in the ultrahigh-dimensional setting, the only thing that needs to be controlled is the tuning parameter $\tau_n$, provided that our hyperpriors in (\ref{mbspmodel}) have the form (\ref{densityxi}). However, in the high-dimensional regime, $p_n$ is allowed to grow at nearly exponential rate, and therefore, the rate of decay for $\tau_n$ from Condition \ref{As:C3} necessarily needs to be much faster. Intuitively, this makes sense because we must sum over $p_n q$ terms in order to compute the Frobenius normed difference in Theorem \ref{Th:4}. 

Taken together, Theorems \ref{Th:3} and \ref{Th:4} both provide theoretical justification for the use of global-local shrinkage priors for multivariate linear regression. Even when we allow the number of predictors to grow at nearly exponential rate, the posterior distribution under MBSP (\ref{mbspmodel}) is able to consistently estimate $\mathbf{B}_0$ in (\ref{B0truth}). Our result is also very general in that a wide class of shrinkage priors, as indicated in Table \ref{table:priors}, can be used for the hyperpriors $\xi_i$'s in (\ref{mbspmodel}). 

\section{Implementation of the MBSP Model} \label{Implementation}

In this section, we demonstrate how to implement the MBSP model using the three parameter beta normal (TPBN) mixture family \cite{ArmaganClydeDunson2011, LibbyNovick1982}. We choose the TPBN family because it is rich enough to generalize several well-known polynomial-tailed priors. Although we focus on the TPBN family, our model can easily be implemented for other global-local shrinkage priors (such as the Student's t prior or the generalized double Pareto prior) using similar techniques as the ones we describe below. 

\subsection{TPBN Family}
A random variable $y$ said to follow the three parameter beta density, denoted as $\mathcal{TPB}(u, a, \tau)$, if
\begin{equation*} 
\pi(y) = \frac{\Gamma (u+a)}{\Gamma(u) \Gamma(a)} \tau^a y^{a-1} (1-y)^{u-1} \left\{ 1-(1-\tau) y \right\}^{-(u+a)}.
\end{equation*}
In univariate regression, a global-local shrinkage prior of the form 
\begin{equation} \label{pixiTPBN}
\begin{array}{l}
\beta_i | \tau, \xi_i  \sim \mathcal{N}(0, \tau \xi_i), \hspace{.2cm} i = 1, \ldots, p, \\
\pi(\xi_i) =  \frac{\Gamma(u+a)}{\Gamma(u) \Gamma(a)} \xi_i^{u-1} (1+\xi_i)^{-(u+a)}, \hspace{.2cm} i = 1, \ldots, p,
\end{array}
\end{equation} 
 may therefore be represented alternatively as
\begin{equation} \label{pixiTPBN2}
\begin{array}{l}
\beta_i | \nu_i  \sim  N(0, \nu_i^{-1} - 1), \\
\nu_i   \sim \mathcal{TPB}(u, a, \tau).
\end{array}
\end{equation}
After integrating out $\nu_i$ in (\ref{pixiTPBN2}), the marginal prior for $\beta_i$ is said to belong to the TPBN family. Special cases of (\ref{pixiTPBN2}) include the horseshoe prior ($u = 0.5, a = 0.5$), the Strawderman-Berger prior ($u=1, a=0.5$), and the normal-exponential-gamma (NEG) prior ($u = 1, a>0$). By Proposition 1 of \citet{ArmaganClydeDunson2011}, (\ref{pixiTPBN}) and (\ref{pixiTPBN2}) can also be written as a hierarchical mixture of two Gamma distributions,
\begin{equation} \label{pixiTPBN3}
\beta_i | \psi_i \sim N(0, \psi_i), \hspace{.3cm} \psi_i | \zeta_i \sim \mathcal{G}(u, \zeta_i), \hspace{.3cm} \zeta_i \sim \mathcal{G}(a, \tau),
\end{equation}
 where $\psi_i = \xi_i \tau$.

\subsection{The MBSP-TPBN Model}

Taking our MBSP model (\ref{mbspmodel}) with the TPBN family as our chosen prior and placing an inverse-Wishart conjugate prior on $\mathbf{\Sigma}$, we can construct a specific variant of the MBSP model which we term the MBSP-TPBN model. For our theoretical study of MBSP, we assumed $\mathbf{\Sigma}$ to be known and the dimension of the responses $q$ to be fixed (and thus, $q < n$ for large $n$). However, in order for our model to be implemented in finite samples, $q$ can be of any size (including $q \gg n$), provided that the posterior distribution is proper. The use of an inverse-Wishart prior ensures posterior propriety.

Reparametrizing the variance terms $\tau \xi_i, 1 \leq i \leq p$, in terms of the $\psi_i$'s from (\ref{pixiTPBN3}), the MBSP-TPBN model is as follows:
\begin{equation}\label{mbspTPBN}
\begin{array}{c}
\mathbf{Y| X, B, \Sigma} 
	\sim \mathcal{MN}_{n \times q} (\mathbf{XB}, \mathbf{I}_n, \mathbf{\Sigma}),  \numbereqn \\ 
	\mathbf{B} | \psi_1, ..., \psi_p, \mathbf{\Sigma}  \sim \mathcal{MN}_{p \times q} (\mathbf{O}, \text{diag}(\psi_1, \ldots, \psi_p), \mathbf{\Sigma}), \\ 
          \psi_i | \zeta_i \overset{\text{ind}}{\sim} \mathcal{G} (u, \zeta_i), i = 1, \ldots, p, \\
          \zeta_i \overset{\text{i.i.d.}}{\sim} \mathcal{G} (a, \tau), i = 1, \ldots, p, \\
\mathbf{\Sigma \sim} \mathcal{IW} (d, k \mathbf{I}_q),
\end{array}
 \end{equation} 
where $u$, $a$, $d$, $k$, and $\tau$ are appropriately chosen hyperparameters. The MBSP-TPBN model can be implemented using the \textsf{R} package \texttt{MBSP}, which is available on the Comprehensive \textsf{R} Archive Network (CRAN).

\subsubsection{Computational Details} \label{ComputationalDetails}
The full conditional densities under model (\ref{mbspTPBN}) are available in closed form, and hence, can be implemented straightforwardly using Gibbs sampling. Moreover, by suitably modifying an algorithm introduced by \citet{BhattacharyaChakrabortyMallick2016} for drawing from the matrix-normal density (\ref{matrix-normal}), we can significantly reduce the computational complexity of sampling from the full conditional density for $\mathbf{B}$ from $O(p^3)$ to $O(n^2 p)$ when $p \gg n$. We provide technical details for our Gibbs sampling algorithm and our algorithm for sampling efficiently from the conditional density for $\mathbf{B}$ in Section 3 of the supplemental materials (see \ref{App:A}).

In our experience, with good initial estimates for $\mathbf{B}$ and $\mathbf{\Sigma}$, $(\mathbf{B}^{(\textrm{init})}, \mathbf{\Sigma}^{(\textrm{init})})$, the Gibbs sampler converges quite quickly, usually within 5000 iterations. In Section 3 of the supplementary materials (see \ref{App:A}), we describe how to initialize $(\mathbf{B}^{(\textrm{init})}, \mathbf{\Sigma}^{(\textrm{init})} )$. In the supplementary materials, we also provide history plots of the draws from the Gibbs sampler for individual coefficients of $\mathbf{B}$ from experiment 5 ($n=100, p =500, q = 3$) and experiment 6 ($n = 150, p = 1000, q=4$) of our simulation studies in Section \ref{SimulationStudies}, which illustrate rapid convergence.

Although our algorithm is efficient, Gibbs sampling can still be prohibitive if $p$ is extremely large (say, on the order of millions). In this case, we recommend first screening the $p$ covariates based on the magnitude of their marginal correlations with the responses $(y_1, \ldots, y_q)$ and then implementing the MBSP model on the reduced subset of covariates. This marginal screening technique for dimension reduction has long been advocated for ultrahigh-dimensional problems, even for non-Bayesian approaches (e.g., \cite{FanLv2008, FanSong2010}). Faster alternatives to MCMC to handle extremely large $p$ are also worth exploring in the future.

\subsubsection{Specification of Hyperparameters $\tau$, $d$, and $k$} \label{hyperparameterspec}

Just as in (\ref{mbspmodel}), the $\tau$ in (\ref{mbspTPBN}) continues to act as a global shrinkage parameter. A natural question is how to specify an appropriate value for $\tau$. \citet{ArmaganClydeDunson2011} recommend setting $\tau$ to the expected level of sparsity. Given our theoretical results in Theorems \ref{Th:3} and \ref{Th:4}, we set $\tau \equiv \tau_n = 1 / (p \sqrt{n \ln n})$. This choice of $\tau$ satisfies the sufficient conditions for posterior consistency in  both the low-dimensional and the high-dimensional settings when $\mathbf{\Sigma}$ is fixed and known.

In order to specify the hyperparameters $d$ and $k$ in the $\mathcal{IW}(d, k \mathbf{I}_q)$ prior for $\mathbf{\Sigma}$, we appeal to the arguments made by \citet{BrownVannucciFearn1998}. As noted by \citet{BrownVannucciFearn1998}, if we set $d = 3$, then $\mathbf{\Sigma}$ has a finite first moment, with $\mathbb{E} (\mathbf{\Sigma}) = k/(d-2) \mathbf{I}_q = k \mathbf{I}_q$. Additionally, as argued in \citet{BhadraMallick2013} and \citet{BrownVannucciFearn1998}, $k$ should \textit{a priori} be comparable in size with the likely variances of $\mathbf{Y}$ given $\mathbf{X}$. Accordingly, we take our initial estimate of $\mathbf{B}$ from the Gibbs sampler, $\mathbf{B}^{(\textrm{init})}$ (specified in Section \ref{ComputationalDetails}), and take $k$ as the variance of the residuals, $\mathbf{Y} - \mathbf{X} \mathbf{B}^{(\textrm{init})}$.

\subsection{Variable Selection} \label{VariableSelection}
Although the MBSP model (\ref{mbspmodel}) and the MBSP-TPBN model (\ref{mbspTPBN}) produce robust estimates for $\mathbf{B}$, they do not produce exact zeros. In order to use model (\ref{mbspTPBN}) for variable selection, we recommend looking at the 95\% credible intervals for each entry $b_{ij}$ in row $i$ and column $j$.  If the credible intervals for every single entry in row $i, 1 \leq i \leq p$, contain zero, then we classify predictor $i$ as an irrelevant predictor. If at least one credible interval in row $i, 1 \leq i \leq p$ does not contain zero, then we classify $i$ as an active predictor. The empirical performance of this variable selection method seems to work well, as shown in Section \ref{SimulationsAndDataAnalysis}.

\section{Simulations and Data Analysis}  \label{SimulationsAndDataAnalysis}

\subsection{Simulation Studies}  \label{SimulationStudies}
For our simulation studies, we implement model (\ref{mbspTPBN}) using our \textsf{R} package \texttt{MBSP}. We specify $u = 0.5, a = 0.5$ so that the polynomial-tailed prior that we utilize is the horseshoe prior. The horseshoe is known to perform well in simulations \cite{CarvahoPolsonScott2010, VanDerPasKleijnVanderVaart2014}. We set $\tau = 1/ (p \sqrt{n \ln n})$, $d = 3$, and $k$ comparable to the size of likely variance of $\mathbf{Y}$ given $\mathbf{X}$.

In all of our simulations, we generate data from the multivariate linear regression model (\ref{Y=XB+E}) as follows. The rows of the design matrix $\mathbf{X}$ are independently generated from $N_p (\mathbf{0}, \bm{\Gamma})$, where $\mathbf{\Gamma} = (\Gamma_{ij})_{p \times p}$ with $\Gamma_{ij} = 0.5^{|i-j|}$. The sparse $p \times q$ matrix $\mathbf{B}$ is generated by first randomly selecting an active set of predictors, $\mathcal{A} \subset \{1, 2, ..., p \}$. For rows with indices in the set $\mathcal{A}$, we independently draw every row element from Unif($[-5, -0.5] \cup [0.5, 5]$). All the other rows in $\mathbf{B}$, i.e. $\mathcal{A}^C$, are then set equal to zero. Finally, the rows of the noise matrix $\mathbf{E}$ are independently generated from $N_q (\mathbf{0}, \mathbf{\Sigma})$, where $\mathbf{\Sigma} = (\Sigma_{ij})_{q \times q}$ with $\Sigma_{ij} = \sigma^2 (0.5)^{|i-j|}, \sigma^2 = 2$.  We consider six different simulation settings with varying levels of sparsity.
\begin{itemize}
\item Experiment 1 ($p < n$): $n = 60, p = 30, q = 3$, 5 active predictors (sparse model).
\item Experiment 2 ($p < n$): $n = 80, p = 60, q = 6$, 40 active predictors (dense model).
\item Experiment 3 ($p > n$): $n = 50, p = 200, q = 5$, 20 active predictors (sparse model).
\item Experiment 4 ($p > n$): $n = 60, p = 100, q = 6$, 40 active predictors (dense model).
\item Experiment 5 ($p \gg n$): $n = 100, p = 500, q = 3$, 10 active predictors (ultra-sparse model).
\item Experiment 6 ($ p \gg n$): $n = 150, p = 1000, q = 4$, 50 active predictors (sparse model).
\end{itemize}
The Gibbs sampler described in Section \ref{ComputationalDetails} is efficient in handling the two $p \gg n$ setups in experiments 5 and 6. Running on an Intel Xeon E5-2698 v3 processor, the Gibbs sampler runs about 761 iterations per minute for Experiment 5 and about 134 iterations per minute for Experiment 6. In all our experiments, we run Gibbs sampling for 15,000 iterations, discarding the first 5000 iterations as burn-in.

As our point estimate for $\mathbf{B}$, we take the posterior median $\widehat{\mathbf{B}} = (\widehat{b}_{ij})_{p \times q}$.  To perform variable selection, we inspect the 95\% individual credible interval for every entry and classify predictors as irrelevant if all of the $q$ intervals in that row contain 0, as described in Section \ref{VariableSelection}. We compute mean squared errors (MSEs) rescaled by a factor of 100, as well as the false discovery rate (FDR), false negative rate (FNR), and overall misclassification probability (MP) as follows:
\begin{equation*}
\begin{array}{rl}
\textrm{MSE}_{\textrm{est}} & = 100 \times || \widehat{\mathbf{B}} - \mathbf{B} ||_F^2 / (pq), \\
\textrm{MSE}_{\textrm{pred}} & = 100 \times || \mathbf{X} \widehat{\mathbf{B}} - \mathbf{X} \mathbf{B} ||_F^2 / (nq), \\
\textrm{FDR} & = \textrm{FP / (TP + FP)}, \\
\textrm{FNR} & = \textrm{FN / (TN + FN)}, \\
\textrm{MP} & = (\textrm{FP + FN}) / (pq),
\end{array}
\end{equation*}
where FP, TP, FN, and TN denote the number of false positives, true positives, false negatives, and true negatives respectively.

We compare the performance of the MBSP-TPBN estimator with that of four other row-sparse estimators of $\mathbf{B}$. An alternative Bayesian approach based on the spike-and-slab formulation is studied. Namely, we consider the multivariate Bayesian group lasso posterior median estimator with a spike-and-slab prior (MBGL-SS), introduced by \citet{LiquetMengersenPettittSutton2017}, which applies a spike-and-slab prior with a point mass $\mathbf{0}^{m_g q}$ for the $g$th group of covariates, which corresponds to $m_g$ rows of $\mathbf{B}$. When the grouping structure of the covariates is not available, we can still utilize the MBGL-SS method by applying the spike-and-slab prior to each individual row of $\mathbf{B}$. In our study, we consider each predictor as its own ``group'' (i.e., $m_g = 1, g = 1, \ldots, p$) so that individual rows are shrunk to $\mathbf{0}_q^\top$.  This method can be implemented in \textsf{R} using the \texttt{MBSGS} package.

In addition, we compare the performance of MBSP-TPBN to three different frequentist point estimators obtained through regularization penalties on the rows of $\mathbf{B}$. In the \textsf{R} package \texttt{glmnet} \cite{FriedmanHastieTibshirani2010}, there is an option to fit the following model to multivariate data, which we call the multivariate lasso (MLASSO) method:
\begin{equation*}
\widehat{\mathbf{B}}^{\textrm{MLASSO}} = \displaystyle \arg \min_{\mathbf{B} \in \mathbb{R}^{p \times q}} \left( || \mathbf{Y}-\mathbf{X B} ||_F^2 + \lambda \displaystyle \sum_{j=1}^{p} || \mathbf{b}_j ||_2 \right).
\end{equation*}
The MLASSO model applies an $\ell_1$ penalty to each of the rows of $\mathbf{B}$ to shrink entire row estimates to be  $\mathbf{0}_q^\top$. We also compare the MBSP-TPBN estimator to the row-sparse reduced-rank regression (SRRR) estimator, introduced by \citet{ChenHuang2012}, which uses an adaptive group lasso penalty on the rows of $\mathbf{B}$, but which further constrains the solution to be rank-deficient. Finally, we compare our method to the sparse partial least squares estimator (SPLS), introduced by \citet{ChunKeles2010}. SPLS combines partial least squares (PLS) regression with a regularization penalty on the rows of $\mathbf{B}$ in order to obtain a row-sparse PLS estimate of $\mathbf{B}$. The SRRR and SPLS methods are available in the \textsf{R} packages \texttt{rrpack} and \texttt{spls}.

Table \ref{Table:2} shows the results averaged across 100 replications for the MBSP-TPBN model (\ref{mbspTPBN}), compared with MBGL-SS, LSGL, and SRRR. As the results illustrate, the Bayesian methods tend to outperform the frequentist ones in the low-dimensional case where $p < n$. In the two low-dimensional experiments (experiments 1 and 2), the MBGL-SS estimator performs the best across all of our performance metrics, with the MBSP-TPBN model following closely behind. 

However, in all the high-dimensional ($p > n$) settings, MBSP-TPBN significantly outperforms all of its competitors. Table \ref{Table:2} shows that the MBSP-TPBN model has a lower $\textrm{MSE}_{\textrm{est}}$ than the other four methods in experiments 3 through 6. In experiments 5 and 6 (the $p \gg n$ scenarios), the $\textrm{MSE}_{\textrm{est}}$ and $\textrm{MSE}_{\textrm{pred}}$ are both much lower for the MBSP-TPBN model than for the other methods. 

Additionally, using the 95\% credible interval technique in Section \ref{VariableSelection} to perform variable selection, the FDR and the overall MP are also consistently low for the MBSP-TPBN model. Even when the true underlying model is not sparse, as in experiments 2 and 4, MBSP performs very well and correctly identifies most of the signals. In both the ultrahigh-dimensional settings we considered in experiments 5 and 6, the other four methods all seem to report high FDR, while the MBSP's FDR remains very small. 

In short, our experimental results show that the MBSP model (\ref{mbsp}) has excellent finite sample performance for both estimation and selection, is robust to non-sparse situations, and scales very well to large $p$ compared to the other methods. In addition to its strong empirical performance, the MBSP model (as well as the MBGL-SS model) provides a vehicle for uncertainty quantification through the posterior credible intervals.

\begin{table}
  \centering
  \medskip
  \begin{tabularx}{\linewidth}{*{6}{p{.14\linewidth}}}
    \multicolumn{6}{l}{Experiment 1: $n = 60, p = 30, q = 3$. 5 active predictors (sparse model).} \\ \toprule
    \textbf{Method} & \textbf{$\textrm{MSE}_{\textrm{est}}$} & \textbf{$\textrm{MSE}_{\textrm{pred}}$} & \textbf{FDR} & \textbf{FNR} & \textbf{MP} \\ \midrule
    MBSP & 1.146 & 24.842 & 0.015 & 0 & 0.003 \\
    MBGL-SS & \textbf{0.718} & \textbf{17.074} & \textbf{0.005} & 0 & \textbf{0.001} \\
   MLASSO & 2.181 & 41.424 & 0.6412 & 0 & 0.335 \\
  SRRR & 1.646 & 29.256 & 0.3270 & 0 & 0.128 \\
 SPLS & 2.428 & 43.879 & 0.1093 & 0.0019 & 0.028 \\  \bottomrule
  \end{tabularx}

  \medskip

  \begin{tabularx}{\linewidth}{*{6}{p{.14\linewidth}}}
    \multicolumn{6}{l}{Experiment 2: $n = 80, p = 60, q = 6$, 40 active predictors (dense model).} \\ \toprule
    \textbf{Method} & \textbf{$\textrm{MSE}_{\textrm{est}}$} & \textbf{$\textrm{MSE}_{\textrm{pred}}$} & \textbf{FDR} & \textbf{FNR} & \textbf{MP} \\ \midrule
    MBSP & 5.617 & 104.88 & 0.0034 & 0 & 0.0023 \\
    MBGL-SS & \textbf{5.202} & \textbf{101.40} & \textbf{0.0007} & 0 & \textbf{0.0005} \\
   MLASSO & 10.478 & 130.90 & 0.3307 & 0 & 0.330 \\
  SRRR & 5.695 & 104.67 & 0.0491 & 0 & 0.038 \\  
   SPLS & 244.136 & 3633.77 & 0.2071 & 0 & 0.223 \\ \bottomrule
  \end{tabularx}

  \medskip

  \begin{tabularx}{\linewidth}{*{6}{p{.14\linewidth}}}
    \multicolumn{6}{l}{Experiment 3: $n = 50, p = 200, q = 5$, 20 active predictors (sparse model).}  \\ \toprule
    \textbf{Method} & \textbf{$\textrm{MSE}_{\textrm{est}}$} & \textbf{$\textrm{MSE}_{\textrm{pred}}$} & \textbf{FDR} & \textbf{FNR} & \textbf{MP} \\ \midrule
    MBSP & \textbf{1.357} & \textbf{117.52} & \textbf{0.0117} & 0 & \textbf{0.0013} \\
    MBGL-SS & 57.25 & 694.81 & 0.858 & 0.02 & 0.619 \\
   MLASSO & 8.400 & 169.026 & 0.7758 & 0 & 0.349 \\
  SRRR & 17.46 & 161.70 & 0.698 & 0 & 0.307 \\
 SPLS & 48.551 & 2006.03 & 0.422 & 0.033 & 0.103 \\  \bottomrule
  \end{tabularx}

  \medskip

  \begin{tabularx}{\linewidth}{*{6}{p{.14\linewidth}}}
    \multicolumn{6}{l}{Experiment 4: $n = 60, p = 100, q = 6$, 40 active predictors (dense model).} \\ \toprule
    \textbf{Method} & \textbf{$\textrm{MSE}_{\textrm{est}}$} & \textbf{$\textrm{MSE}_{\textrm{pred}}$} & \textbf{FDR} & \textbf{FNR} & \textbf{MP} \\ \midrule
    MBSP & \textbf{11.030} & \textbf{172.89} & \textbf{0.0266} & 0 & \textbf{0.0114} \\
    MBGL-SS & 204.33 & 318.80 & 0.505 & 0.1265 & 0.415 \\
   LSGL & 44.635 & 188.81 & 0.544 & 0 & 0.479 \\
  SRRR & 242.67 & 193.64 & 0.594 & 0 & 0.587 \\
 SPLS & 213.19 & 3909.07 & 0.135 & 0.0005 & 0.005  \\  \bottomrule
  \end{tabularx}

 \medskip

  \begin{tabularx}{\linewidth}{*{6}{p{.14\linewidth}}}
    \multicolumn{6}{l}{Experiment 5: $n = 100, p = 500, q = 3$, 10 active predictors (ultra-sparse model).} \\ \toprule
    \textbf{Method} & \textbf{$\textrm{MSE}_{\textrm{est}}$} & \textbf{$\textrm{MSE}_{\textrm{pred}}$} & \textbf{FDR} & \textbf{FNR} & \textbf{MP} \\ \midrule
    MBSP & \textbf{0.0374} & \textbf{12.888} & \textbf{0.064} & 0 & \textbf{0.0015} \\
    MBGL-SS & 1.327 & 155.51 & 0.483 & 0.0005 & 0.092 \\
   MLASSO & 0.2357 & 75.961 & 0.837 & 0 & 0.115 \\
  SRRR & 0.9841 & 49.428 & 0.688 & 0 & 0.104 \\  
   SPLS & 0.3886 & 138.62 & 0.1355 & 0.0005 & 0.005 \\ \bottomrule
  \end{tabularx}

  \begin{tabularx}{\linewidth}{*{6}{p{.14\linewidth}}}
    \multicolumn{6}{l}{Experiment 6: $n = 150, p = 1000, q = 4$, 50 active predictors (sparse model).} \\ \toprule
    \textbf{Method} & \textbf{$\textrm{MSE}_{\textrm{est}}$} & \textbf{$\textrm{MSE}_{\textrm{pred}}$} & \textbf{FDR} & \textbf{FNR} & \textbf{MP} \\ \midrule
    MBSP & \textbf{0.0155} & \textbf{8.934} & \textbf{0.0025} & 0.00003 & \textbf{0.00016} \\
    MBGL-SS & 1.327 & 155.51 & 0.483 & 0.0005 & 0.092 \\
   MLASSO & 1.982 & 181.95 & 0.810 & 0 & 0.214 \\
  SRRR & 0.9841 & 49.428 & 0.688 & 0 & 0.104 \\
 SPLS & 25.560 & 8631.92 & 0.420 & 0.021 & 0.051  \\ \bottomrule
  \end{tabularx}

  \caption{Simulation results for MBSP-TPBN, compared with MBGL-SS, MLASSO, SRRR, and SPLS, averaged across 100 replications. }
  \label{Table:2}
\end{table}

\subsection{Yeast cell cycle data analysis}  \label{DataAnalysisExample}
We illustrate the MBSP methodology on a yeast cell cycle data set. This data set was first analyzed by \citet{ChunKeles2010} and is available in the \texttt{spls} package in \texttt{R}. Transcription factors (TFs) are sequence-specific DNA binding proteins which regulate the transcription of genes from DNA to mRNA by binding specific DNA sequences. In order to understand their role as a regulatory mechanism, one often wishes to study the relationship between TFs and their target genes at different time points.  In this yeast cell cycle data set, mRNA levels are measured at 18 time points seven minutes apart (every 7 minutes for a duration of 119 minutes). The $542 \times 18$ response matrix $\mathbf{Y}$ consists of 542 cell-cycle-regulated genes from an $\alpha$ factor arrested method, with columns corresponding to the mRNA levels at the 18 distinct time points. The $542 \times 106$ design matrix $\mathbf{X}$ consists of the binding information of a total of 106 TFs. 

In practice, many of the TFs are not actually related to the genes, so our aim is to recover a parsimonious model with only a tiny number of the truly statistically significant TFs. To perform variable selection, we fit the MBSP-TPBN model (\ref{mbspTPBN}) and then use the 95\% credible interval method described in Section \ref{VariableSelection}. Beyond identifying significant TFs, we assess the predictive performance of the MBSP-TPBN model (\ref{mbspTPBN}) by performing five-fold cross validation, using 80 percent of the data as our training set to obtain an estimate of $\mathbf{B}$, $\widehat{\mathbf{B}}^{\textrm{train}}$. We take the posterior median as $\widehat{\mathbf{B}}^{\textrm{train}} = (\widehat{b}_{ij})^{\textrm{train}}$ and use it to compute the mean squared error of the residuals on the remaining 20 percent of the left-out data. We repeat this five times, using different training and test sets each time, and take the average MSE as our mean squared predictor error (MSPE). To make our analysis more clear, we scale the MSPE by a factor of 100.

Table \ref{Table:3} shows our results compared with the MBGL-SS, MLASSO, SRRR, and SPLS methods. MBSP-TPBN selects 12 of the 106 TFs as significant, so we do recover a parsimonious model. All five methods selected the TFs, ACE2, SWI5, and SWI6. The two Bayesian methods seem to recover a much more sparse model than the frequentist methods. In particular, the MLASSO method has lowest MSPE, but it selects 78 of the 106 TFs as significant, suggesting that there may be overfitting in spite of the regularization penalty on the rows of $\mathbf{B}$. Our results suggest that the frequentist methods may have good predictive performance on this particular data set, but at the expense of parsimony. In practice, sparse models are preferred for the sake of interpretability, and our numerical results illustrate that the MBSP model recovers a sparse model with competitive predictive performance.

\begin{table}[t!]
  \centering
\begin{tabular}{l*{3}{c}r}
\hline
\textbf{Method}  & \textbf{Number of Proteins Selected} & \textbf{MSPE} \\
\hline
MBSP & 12 & 18.673 \\
MBGL-SS & 7 & 20.093 \\
MLASSO & 78 & 17.912 \\
SRRR & 44 & 18.204 \\
SPLS & 44 & 18.904 \\
\hline
\end{tabular}
  \caption{Results for analysis of the yeast cell cycle data set. The MSPE has been scaled by a factor of 100. In particular, all fives models selected the three TFs, ACE2, SWI5, and SWI6 as significant.}
  \label{Table:3}
\end{table}

\begin{figure}[t!]
\centering
\includegraphics[scale=0.8]{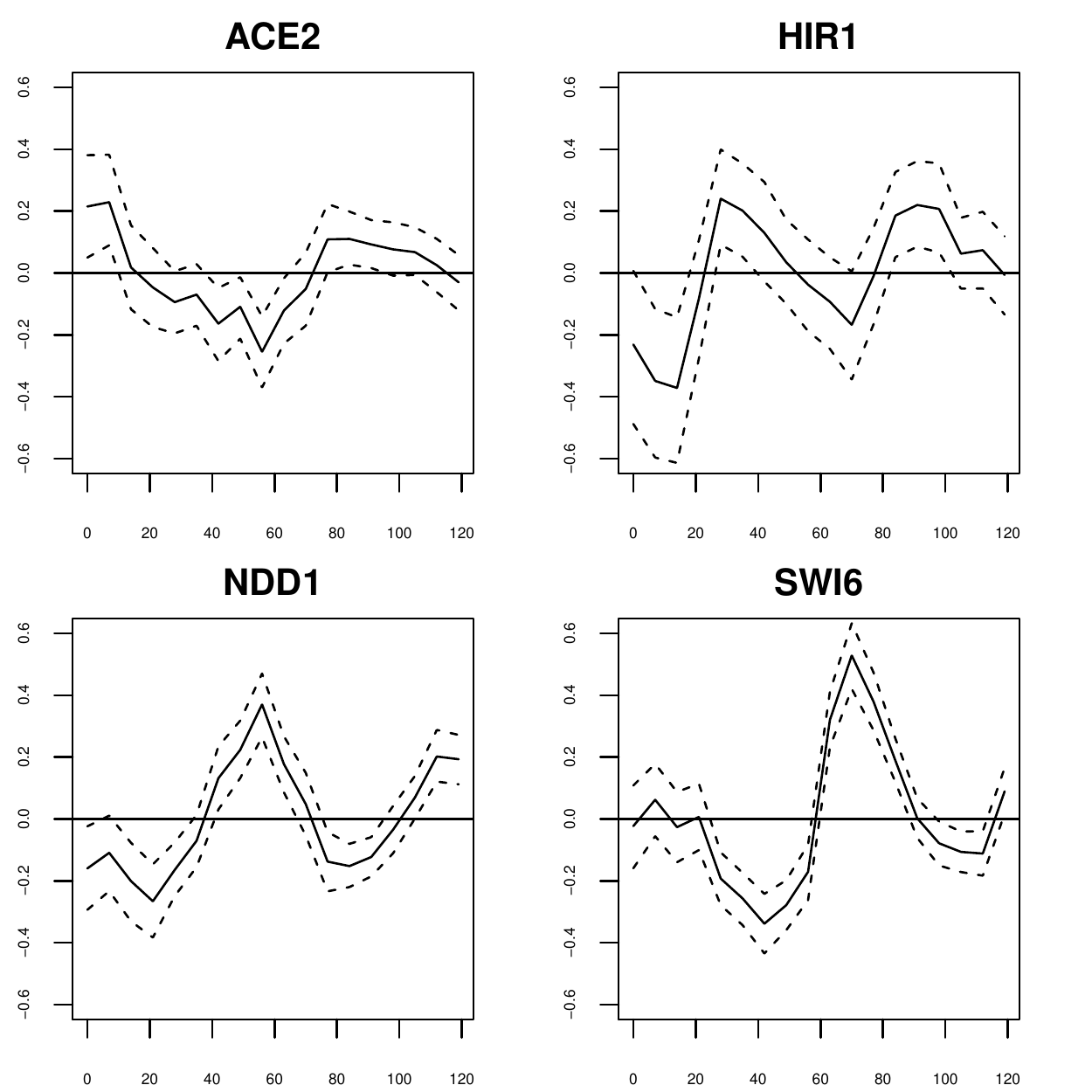} 
\caption{Plots of the estimates and 95\% credible bands for four of the 10 TFs that were deemed as significant by the MBSP-TPBN model. The x-axis indicates time (minutes) and the y-axis indicates the estimated coefficients.}
\label{fig:1}
\end{figure}

Finally, Figure \ref{fig:1} illustrates the posterior median estimates and the 95\% credible bands for four of the 10 TFs that were selected as significant by the MBSP-TPBN model. These plots illustrate that the standard errors under the MBSP-TPBN model are not too large. One of the potential drawbacks of using credible intervals for selection is that these intervals may be too conservative, but we see that it is not the case here. This plot, combined with our earlier simulation results and our data analysis results, provide empirical evidence for using the MBSP model for estimation and variable selection.  However, further theoretical investigation is warranted in order to justify the use of marginal credible intervals for variable selection. In particular, \citet{VanDerPasEtAl2017} showed that marginal credible intervals may provide overconfident uncertainty statements for certain large signal values when applied to estimating normal mean vectors, and the same issue could be present here.

\section{Conclusion and Future Work} \label{Conclusion}

In this paper, we have introduced a method for sparse multivariate Bayesian estimation with shrinkage priors (MBSP). Previously, global-local shrinkage priors have mainly been used in univariate regression or in the estimation of normal mean vectors. Our paper extends their use to the multivariate linear regression framework. 

Our paper makes several important contributions to methodology and theory. First, our model may be used for sparse multivariate estimation for $p, n,$ and $q$ of any size. To motivate the MBSP model, we have shown that the posterior distribution can consistently estimate $\mathbf{B}$ in (\ref{Y=XB+E}) in both the low-dimensional and ultrahigh-dimensional settings where $p$ is allowed to grow nearly exponentially with $n$ (with the response dimension $q$ fixed). This appears to be the first paper to provide sufficient conditions for ultrahigh-dimensional posterior consistency under model (\ref{Y=XB+E}) in the statistical literature. Moreover, our method is general enough to encompass a large family of heavy-tailed priors, including the Student's-t prior, the horseshoe prior, the generalized double Pareto prior, and others.

The MBSP model (\ref{mbspmodel}) can be implemented using straightforward Gibbs sampling. We implemented a fully Bayesian version of it with an appropriate prior on $\mathbf{\Sigma}$ and with polynomial-tailed priors belonging to the TPBN family, using the horseshoe prior as a special case. By examining the 95\% posterior credible intervals for every element in each row of the posterior conditional distribution of $\mathbf{B}$, we also showed how one could use the MBSP model for variable selection. Through simulations and data analysis on a real data set, we have illustrated that our model has excellent performance in finite samples for both estimation and variable selection.

\subsection{Future Work}

Although our paper addresses a long-standing gap between theory and application for Bayesian multivariate linear regression, much still remains unknown. In this paper, we demonstrated that the MBSP model (\ref{mbspmodel}) could achieve posterior consistency in both low-dimensional ($p = o(n)$) and ultrahigh-dimensional ($\ln p = o(n)$) settings. The next step is to quantify the posterior contraction rate. In the present context of multivariate linear regression, we say that the posterior distribution contracts at the rate $r_n$ if
\begin{equation*}
\Pi_n ( || \mathbf{B}_n - \mathbf{B}_0 ||_F > M_n r_n | \mathbf{Y}_n ) \rightarrow 0 \textrm{ a.s. } \mathbb{P}_0 \textrm{ as } n \rightarrow \infty,
\end{equation*}
for every $M_n \rightarrow \infty$ as $n \rightarrow \infty$. In the context of high-dimensional \textit{univariate} regression, several authors (e.g., \cite{CastilloSchmidtHieberVanderVaart2015}, \cite{RockovaGeorge2016}) have attained optimal posterior contraction rates of $O( \sqrt{s \ln (p) / n})$ with respect to the $\ell_1$ and $\ell_2$ norms (where $s$ denotes the number of active predictors). It is worth noting that $ \sqrt{s \ln (p) /n}$ is the familiar minimax rate of convergence under squared error loss for a number of frequentist point estimators, including the Dantzig selector \cite{CandesTao2007}, the scaled lasso \cite{SunZhang2012}, and the LASSO \cite{Tibshirani1996}. We conjecture that under suitable regularity conditions and compatibility conditions on the design matrix, the MBSP model can attain a similarly optimal posterior rate of contraction.

Additionally, we could investigate if posterior consistency and optimal posterior contraction rates can be achieved if we allow the number of response variables $q$ to diverge to infinity in the MBSP model. From an implementation standpoint, $q$ can be of any size, but for our theoretical investigation of the MBSP model, we assumed $q$ to be fixed. If $q$ is allowed to grow as sample size grows, then some sort of sparsity assumption for the response variables may need to be imposed. We surmise that novel techniques would also be needed to prove posterior consistency in this scenario, since the distributional theory we used to prove our consistency results may not apply if $q$ is no longer fixed.  

Extension of our posterior consistency results to the case where $\mathbf{\Sigma}$ is unknown and endowed with a prior also remains an open problem. In this case, we need to integrate out $\mathbf{\Sigma}$ in order to work with the marginal density of the prior on $\mathbf{B}$. If we assume the standard inverse-Wishart prior on $\mathbf{\Sigma}$, this gives rise to a matrix-variate t distribution. Handling this density is very nontrivial and would require significantly different techniques than the ones we used to establish posterior consistency in Section \ref{mbspconsistent}. Nevertheless, this warrants future investigation.

For variable selection with the MBSP model, we relied on the post hoc method of examining the 95\% credible intervals for each entry of the estimated coefficients matrix for $\mathbf{B}$. Further theoretical justification for this selection method is needed. Other possible thresholding rules should also be investigated. Because scale-mixture shrinkage priors place zero probability at exactly zero, we must necessarily use thresholding to perform variable selection. How to optimally choose this threshold (or thresholds) in high-dimensional settings remains an active area of research. 

All the aforementioned are very important open problems in Bayesian multivariate linear regression, and we hope that the methodology and theory introduced in this paper can serve as the foundation for further developments in this area.

\section*{Acknowledgments}
The authors are grateful to the Associate Editor and two referees for their helpful comments on an earlier version of this paper.

\begin{appendix}

\section{Supplementary Material} \label{App:A}
Supplementary material related to this article can be found online at 

\href{https://people.clas.ufl.edu/raybai07/files/MBSP-Supplement.pdf}{\tt https://people.clas.ufl.edu/raybai07/files/MBSP-Supplement.pdf}.

\end{appendix}

\section*{References}

\bibliography{Reference}

\end{document}